\documentclass[epj]{svjour}
\usepackage{graphics}
\usepackage{amssymb}
\usepackage{amsmath}
\newcommand{\half}{\mbox{${\textstyle \frac{1}{2}}$}}           
\newcommand{\sixth}{\mbox{${\textstyle \frac{1}{6}}$}}          
\newcommand{\fmn}[2]{\mbox{${\textstyle \frac{#1}{#2}}$}}
\newcommand{\rd}{\mbox{{\rm d}}}
\newcommand{\bmath}[1]{\mbox{\boldmath $#1$}}
\newcommand{\fsi}{\emph{fsi}}
\newcommand{\pde}{$p\,d\rightarrow p\,d\,\eta$}
\newcommand{\pdhe}{$p\,d\to \,^{3}$He$\,\eta$}

\begin{document}
\title{Theoretical Description of the \pde\ Reaction Near Threshold}

\author{Ulla Tengblad\inst{1} \and G\"{o}ran F\"{a}ldt\inst{1}
\and Colin Wilkin\inst{2}}

\institute{ Department of Radiation Sciences, Box 535, S--751 21
Uppsala, Sweden, \email{goran.faldt@tsl.uu.se} \and Department of
Physics \& Astronomy, UCL, London WC1E 6BT, UK,
\email{cw@hep.ucl.ac.uk}}

\date{Received: \today / Revised version: date}

\abstract{The contributions of three different types of driving
terms are included in the estimation of the \pde\ reaction at low
energies. Near threshold, it is predicted that a two--step model
involving an intermediate pion should be the most important but,
as the energy approaches the threshold for $\eta$ production in
the free nucleon--nucleon reaction, a pick--up mechanism with a
spectator proton should become dominant. The total cross sections
are underestimated by about a factor of two compared to
experimental data but the discrepancies in the angular
distributions are more serious, with no sign in the data for the
peaks corresponding to the pick--up diagram.
 \PACS{{25.10.+s} {Nuclear reactions involving few--nucleon systems}\and
 {13.60.Le} {Meson production} \and {14.40.Aq} {pi, K, and eta mesons}}}
\maketitle
%
%
\section{Introduction}

The great interest in the production of heavy mesons near
threshold started with measurements of the \pdhe\ reaction, which
showed a surprisingly strong cross section with an anomalous
energy dependence~\cite{Berger}. An impulse approximation
description of the process, where the production takes place in
nucleon--nucleon scattering, with the other nucleon in the
deuteron being essentially a spectator, greatly underestimates the
observed rate because of the high momentum components required in
the nuclear wave functions~\cite{Germond}. To share the large
momentum transfer between the nucleons, a mechanism was proposed
whereby a pion was produced in an initial $pp\to d\pi^+$ reaction
to be followed by the production of the observed meson through a
secondary $\pi^+n\to\eta p$ process~\cite{Kilian}. Near threshold,
the kinematics are favourable for the final proton and deuteron to
\emph{stick} to form the observed $^3$He. A quantum--mechanical
evaluation of this suggestion~\cite{FW} led to a cross section
that was only about a factor of two lower than a precise
measurement of the reaction rate~\cite{Mayer}. To understand the
situation further, it would be helpful to look at other related
final states.

The total cross section for the \pde\ reaction, where a final
$^3$He is not formed, was measured at two energies very close to
threshold at Saclay~\cite{Hibou}. A preliminary evaluation of the
two--step model with an intermediate pion, that was successful in
the description of the $^3$He$\,\eta$ final state, gave very
promising results when compared to these data~\cite{Ulla}.
Differential as well as total cross sections for this reaction at
higher energies have recently become available from
Uppsala~\cite{Jozef2}, to complement those obtained by the same
group for the \pdhe\ reaction~\cite{Jozef1}. The time therefore
seems opportune to make a further theoretical investigation of the
$p\,d\,\eta$ final state.

The kinematics of the Saclay experiment~\cite{Hibou} are very
close to those where a $^3$He emerges and so it is not surprising
that in this region the two--step model describes the process
well, as it did for the $^3$He$\,\eta$ final state~\cite{FW}. On
the other hand, as one approaches the threshold for $\eta$
production in free nucleon--nucleon collisions, one expects a
pick--up diagram, corresponding to a quasi--free $pn\to d\,\eta$
production on the neutron in the target, to dominate. Between
these two extremes there is also the possibility of a contribution
from the impulse approximation diagram with a quasi--free $pN\to
pN\,\eta$ on a proton or neutron that is bound both initially and
finally in the deuteron. Away from threshold, these terms may not
be universally suppressed, as they are for $^3$He$\,\eta$, because
of the greater flexibility in the $p\,d\,\eta$ kinematics; in
certain parts of phase space, the momentum transfers between
initial and final deuterons are minimised.

The three types of contributions are described further in sect.~2,
where the general kinematics are discussed. The following three
sections are devoted to the evaluation of the cross section
distributions for the individual terms. Uncertainties in the
phases of the amplitudes leads us to neglecting interferences.
This may not be too dangerous because the various models tend to
populate different parts of what is a three--body phase space.
More worrying is that we neglect also all final--state
interactions \fsi. Now, unlike the $\eta\,^3$He case~\cite{Mayer},
there is no simple prescription for the inclusion of such effects
when there is an \fsi\ between more than one pair of particles.
Experimentally a threshold enhancement is seen only in the
$\eta\,d$ invariant mass distribution, whereas that for $p\,d$
looks like phase space, despite the presence of the strong
interaction which could lead to the formation of the
$^3$He~\cite{Jozef2}. The results for the total cross sections,
and both angular and invariant mass distributions, are compared
with experiment in sect.~6. The overall production rate is
underestimated by about a factor of two, which is rather similar
to that found for the \pdhe\ reaction within a similar theoretical
approach. However, the angular distributions of both the proton
and deuteron show discrepancies with respect to the data, with the
latter showing no signs of the forward deuteron and backward
proton peaks corresponding to the spectator proton of the pick--up
mechanism. Our conclusions are drawn in sect.~7.
%
%
\section{Defining the problem}
\setcounter{equation}{0}

\begin{figure}[ht]
\begin{center}
\resizebox{0.45\textwidth}{!}{
  \includegraphics{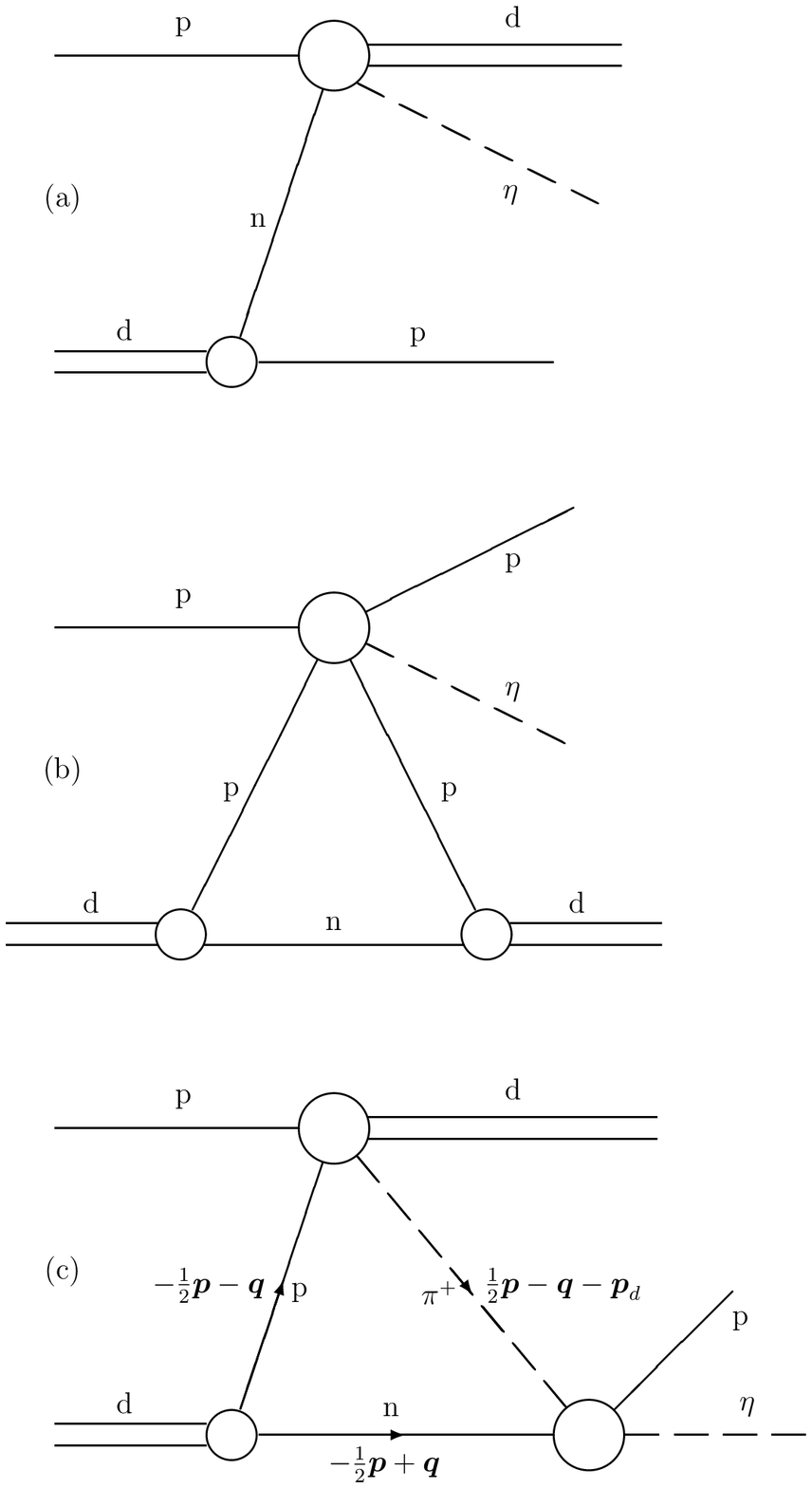}}
 \caption{Three classes of diagrams relevant for the \pde\ reaction.
 (a) The pick--up term which dominates above the $NN$ threshold.
 (b) The impulse (triangle) diagram. There is a similar contribution with
 the internal proton and neutron interchanged.
 (c) The two--step model with an intermediate pion. The contribution where
 the $\pi^+$ is replaced by a $\pi^0$ is related to this through isospin
 invariance.}
 \label{fig1}
\end{center}
\end{figure}

The three classes of diagram that we evaluate for the \pde\
reaction are illustrated in Fig.~\ref{fig1}. It is generally
assumed that the neutron--exchange diagram (a) dominates the
reaction above the free $NN$ threshold and, indeed, it is this
hypothesis that is the basis of the extraction of the quasi--free
$pn\to d\,\eta$ cross section on a moving neutron from \pde\
data~\cite{Stina}. Below threshold, the presence of a spectator
proton in this case will bias the distributions to low $d\,\eta$
invariant masses.

The impulse approximation diagram of Fig.~\ref{fig1}b is often
used to model coherent reactions on the deuteron at high energies,
where the momentum transfer to the final deuteron can be small.
There is, of course, a second term where the $\eta$ production
takes place on the neutron and some theoretical model is required
to deduce the relative phases of the spin--isospin input
amplitudes needed to make reliable estimates here~\cite{FW2}. A
final--state interaction between the proton and deuteron to form
an $^3$He for either the diagrams in Fig.~\ref{fig1}a,b would lead
to the triangle graph, whose contribution has been shown to be
small for \pdhe~\cite{Germond}. That they are not necessarily
negligible here is due to the fact that the final $dp$ pair does
not have to emerge with low excitation energy.

The differential cross section for the \pde\ reaction is
determined by the matrix element $\mathcal{M}$ through
\begin{equation}
\label{PSG}
\rd\sigma=\frac{1}{4p\sqrt{s}}\left\{\!\frac{1}{6}\sum_{\textrm{\tiny
spins}}
\left|\mathcal{M}\right|^2\!\right\}(2\pi)^4\delta^4(P_i-P_f)\prod_{j=1}^{3}
\frac{\rd^3p_j}{(2\pi)^{3}2E_j}\,,
\end{equation}
where the sum is over the spin projections in the initial and
final states. Here $P_i^2=P_f^2=s$ is the square of the total cm
energy and $(E_j,\bmath{p}_j)$ are the energy and three--momentum
of the reaction products. The incident flux factor in the centre
of mass involves the initial proton cm momentum \bmath{p}.

Experimental distributions have been presented in terms of the
angles of the final particles and the invariant masses of pairs of
such particles~\cite{Jozef2}. For this purpose it is convenient to
reduce the phase--space factors to:
\begin{equation}
\label{PSG2}
\rd\sigma=\frac{1}{4p\,s}\left\{\!\frac{1}{6}\sum_{\textrm{\tiny
spins}} \left|\mathcal{M}\right|^2\!\right\}\frac{1}{64\pi^5}\,
p_d\,\rd\Omega_d\,p_{\eta}^*\rd\Omega_{\eta}^*\,\rd m_{\eta p} \,,
\end{equation}
where the final deuteron energy and momentum $(E_d,\bmath{p}_d)$,
as well as the angles $\Omega_d$, are evaluated in the overall cm
system while those of the $\eta$ are evaluated in the $\eta p$
rest frame, where the invariant mass is given by
\begin{equation}
m_{\eta p}^2 = s+m_d^2-2\sqrt{s}\,E_d\:.
\end{equation}
Analogous formulae follow immediately for the other two
combinations of variables.

%
%
\section{The pick--up contribution}
\setcounter{equation}{0}%

Though there is an enhancement at threshold, the energy dependence
of the $pn\to d\,\eta$ total cross section is broadly consistent
with $s$--wave production up to an excess energy of at least
60~MeV~\cite{Stina}. At threshold there is only one cm amplitude,
which can be written as
\begin{equation}
{\cal M}(pn\rightarrow d\eta)= u_{n_c}^{\dagger}
  \left[  G\sqrt{s_{pn}}  \, \bmath{p}_n\cdot\bmath{\epsilon}_d
  \right] u_{p}  \ ,
\label{Eta-threshold-amplitude}
\end{equation}
where $\bmath{p}_n$ is the initial momentum and the cm energy
$\sqrt{s_{pn}}$ arises from the reduction from a relativistic
form. Here $u_{p}$ and $u_{n_c}$ are proton and charge--conjugate
neutron Pauli spinors respectively.

In terms of the amplitude $G$, the total $pn\to d\,\eta$ cross
section is
\begin{equation}
\sigma_T(pn\!\to\! d\eta)= \frac{1}{8\pi} p_{\eta}^*p_p^*\, |G|^2
\:, \label{Eta-threshold}
\end{equation}
where the $p^*$ are evaluated in the $pn$ cm frame. The
experimental data are then consistent with a value of
$|G|^2=(0.046\pm0.011)$~fm$^4$~\cite{Stina}.

The spin--averaged amplitude squared for the pick--up term of
Fig.~\ref{fig1}a reduces to:
\begin{eqnarray}
\nonumber
&&\sixth|{\cal M}(pd\rightarrow pd\eta)|^2\\
&&=\half [ (2\pi)^3 2m_d] (4E_pp_p)^2\,|G|^2\,
 \left\{ \tilde{\varphi}_S(q)^2 +
 \tilde{\varphi}_D(q)^2\right\}\phantom{aaaaaa}
\label{Summary}
\end{eqnarray}
Here the deuteron wave functions $\tilde{\varphi}_S(q)$ and
$\tilde{\varphi}_D(q)$ are normalised
by%
\begin{equation}
\int_0^{\infty}q^2\,\left\{ \tilde{\varphi}_S(q)^2 +
 \tilde{\varphi}_D(q)^2\right\}=1\:.
\end{equation}
They are evaluated at a momentum--squared of
\begin{equation}
\bmath{q}^2 =\frac{m_d^2}{E_{di}^2} (-\half p -p_{p\parallel})^2
  + \bmath{p}_{p\perp}^2 \:,
\end{equation}
where the Lorentz boost has been approximated in the low Fermi
momentum limit and $(p_{p\parallel},\bmath{p}_{p\perp})$ are the
components of the final proton momentum $\bmath{p}_{p}$ parallel
and perpendicular to the momentum $\bmath{p}$ of the incident
deuteron, which has energy $E_{di}$.

%
%
\section{The triangle diagram}
\setcounter{equation}{0}%
The triangle diagram of Fig.~\ref{fig1}b requires as input the
amplitudes for the sub--reactions $NN\rightarrow NN\eta$. The
threshold cm amplitudes for $I=1$ and $I=0$ are respectively:
\begin{eqnarray}
{\cal M}_1(NN\to NN\eta)&=&  \left[\:{\cal W}_{1,s}\:
  {\eta}_f^{\,\dagger}\:\hat{p}\cdot \bmath{\epsilon}_i
  \:\right]\,\bmath{\chi}_f^{\,\,\dagger}\cdot\bmath{\chi}_i\:,
\label{L_1}\\
{\cal M}_0(NN\to NN\eta)&=& \left[\:{\cal W}_{0,t}\:
   \hat{p} \cdot \bmath{\epsilon}_f^{\,\,\dagger}\:{\eta}_i\:
    \right]\,\phi_f^{\,\dagger}\,\phi_i \:,
\label{L_2}
\end{eqnarray}
where $\bmath{\chi}$ and $\phi$ represent the isospin--1 and
isospin--0 configurations of the $NN$ states, with
$\bmath{\epsilon}$ and ${\eta}$ corresponding to the spin--1 and
spin--0 combinations.

The spin--averaged total $\eta$ production cross sections are
\begin{eqnarray}
\nonumber \sigma(pp\!\to\! pp\eta)&=&
\frac{1}{512\pi^2p_{NN}s_{NN}} \frac{m_N}
{\left(1+2m_N/m_{\eta}\right)^{1/2}}\\
\nonumber &&\times Q_{NN\eta}^2\,|{\cal W}_{1,s}|^2\:,\\
\nonumber \sigma(pn\!\to\! pn\eta)&=&
\frac{1}{1024\pi^2p_{NN}s_{NN}} \frac{m_N}
{\left(1+2m_N/m_{\eta}\right)^{1/2}}\\
&&Q_{NN\eta}^2\,\times \left[|\mathcal{W}_{1,s}|^2
+|\mathcal{W}_{0,t}|^2\right] \:, \label{sigmaNN}
\end{eqnarray}
where $p_{NN}$ and $s_{NN}$ are the momentum and square of the
total energy in the cm system, $m_N$ the nucleon mass, and
$Q_{NN\eta}$ the excess energy in the final system~\cite{FW2}.

In terms of an initial deuteron momentum $-\bmath{p}$, a final
deuteron momentum $\bmath{p}_d$, and a momentum transfer
$\bmath{Q}=\bmath{p}_d +\bmath{p}$, the triangle graph is
evaluated in the standard way by putting the \textit{spectator}
nucleon on--shell at an average energy $E_N(\half\bmath{p})$. The
spin--averaged matrix element becomes
\begin{eqnarray}
\nonumber &&\frac{1}{6}\sum_{\text spin} \mid \mathcal{M}(pd\to
pd\eta)\mid^2=\frac{1}{64}\left(\frac{m_d}{4E_N(\bmath{p}/2)}\right)^{\!2}\times\\
\nonumber &&\left[|3W_{1,s}+W_{0,t}|^2\,\left[S_S^2(\half
Q)+S_Q^2(\half Q) +\fmn{2}{3}S_A^2(\half
Q)\right]\right.\\
&&\vphantom{\int}\left.+|3W_{1,s}
-W_{0,t}|^2\,\fmn{4}{3}S_M^2(\half Q)\right]\:.
\end{eqnarray}

The spherical, quadrupole, magnetic and axial form factors
appearing here are defined by
\begin{eqnarray}
\nonumber%
S_S(Q)&=&S_a(Q)+S_b(Q)\,,\\
\nonumber%
S_Q(Q)&=&2S_c(Q)-S_d(Q)/\sqrt{2}\,\\
\nonumber%
S_M(Q)&=&S_a(Q)-\half S_b(Q)+S_c(Q)/\sqrt{2}+\half S_d(Q)\,,\\
S_A(Q)&=&S_a(Q)-\half S_b(Q)-\sqrt{2}S_c(Q)- S_d(Q)\,,
\end{eqnarray}
where the terms on the right hand side represent integrals over
the reduced deuteron wave functions $u(r)$ and $w(r)$:
\begin{eqnarray}
\nonumber%
S_a(Q)&=&\int_0^{\infty}j_0(Qr)\,\left[u(r)\right]^2\,\rd r\,,\\
\nonumber%
S_b(Q)&=&\int_0^{\infty}j_0(Qr)\,\left[w(r)\right]^2\,\rd r\,,\\
\nonumber%
S_c(Q)&=&\int_0^{\infty}j_2(Qr)\,u(r)w(r)\,\rd r\,,\\
S_d(Q)&=&\int_0^{\infty}j_2(Qr)\,\left[w(r)\right]^2\,\rd r\,.
\end{eqnarray}

Experimental data on $\eta$ production in $pp$~\cite{Calen} and
$pn$~\cite{Calen3} collisions above the \fsi\ regions show that
$|W_{1,s}|^2\approx 1.9\times 10^8~\mu$b and $|W_{0,t}|^2\approx
23\times 10^8~\mu$b, though the relative phase of these two
amplitudes is not determined. In a meson-exchange approach, it is
suggested that $\rho$--exchange is the largest term~\cite{FW2}. If
further we take the two amplitudes to be in phase then we find
that
\begin{eqnarray}%
\nonumber |3\mathcal{W}_{1,s}+\mathcal{W}_{0,t}|^2&=&80\times
10^8\:\mu\textrm{b}\:,\\
|3\mathcal{W}_{1,s}-\mathcal{W}_{0,t}|^2&=&0.4\times
10^8\:\mu\textrm{b}\:.
\end{eqnarray}

%
%
\section{The two--step model}
\setcounter{equation}{0}%

The contribution of the two--step model with an intermediate pion
can be estimated using the same techniques as for the more
complicated \pdhe\ reaction, where there are two integration loops
over the unobserved Fermi momenta~\cite{FW}. In terms of the
internal momenta marked in Fig.~1c, the \pde\ matrix element
becomes
\begin{eqnarray}
\nonumber%
\mathcal{M}=\frac{\sqrt{2m_d}}{(2\pi)^{3/2}}\int\rd^3q'\,
\sum_{\textrm{\tiny int}} \mathcal{M}(\pi^+n\to\eta\,p)\times\\
\frac{-i}{2E_{\pi}(\sqrt{s}-E_{\textrm{\tiny int}}+i\varepsilon)}
\mathcal{M}(pp\to \pi^+d)\,\tilde{\Psi}(\bmath{q}')\:,
\label{master}
\end{eqnarray}
where the sum runs over the internal spin indices.

The pion propagator between the matrix elements for the production
and conversion of the pion in eq.~(\ref{master}) has been
approximated by its positive energy pole. The difference between
the external and internal energies, $\Delta E
=\sqrt{s}-E_{\textrm{\tiny int}}$, depends upon the Fermi momentum
\bmath{q} and, following ref.~\cite{FW}, we retain only linear
terms in this integration variable:
\begin{eqnarray}
\nonumber%
\Delta E &=&\sqrt{s}-E_d(\bmath{p}_d) -
E_n(-\half\bmath{p}+\bmath{q})-
E_{\pi}(\half\bmath{p}-\bmath{q}-\bmath{p}_d)\\
&\approx& \Delta E_0 +\bmath{V}\cdot\bmath{q}\:,
\end{eqnarray}
where the mean energy defect
\begin{equation}
\Delta E_0 =\sqrt{s}-E_d(\bmath{p}_d) - E_n(-\half\bmath{p})-
E_{\pi}(\half\bmath{p}-\bmath{p}_d)
\end{equation}
and the relativistic relative velocity between the pion and
neutron
\begin{equation}
\bmath{V}=\bmath{v}_{\pi}-\bmath{v}_{n}=
\left(\frac{E_n+E_{\pi}}{2E_nE_{\pi}}\right)\,\bmath{p}-
\frac{1}{E_{\pi}}\,\bmath{p}_d
\end{equation}
do not depend on the proton or $\eta$ momenta in the final state.

Both the argument of the deuteron momentum space wave function
$\tilde{\Psi}(\bmath{q}')$ and the integration variable are the
Fermi momentum $\bmath{q}'$, Lorentz contracted in the beam
direction; $\bmath{q}'_{\bot}=\bmath{q}_{\bot}$,
$\bmath{q}'_{\|}=\bmath{q}_{\|}/\gamma$, with $\gamma=E_d(p)/m_d
\approx E_n(-\half p)/m_n$.

The linearisation is valid only if relatively small Fermi momenta
are required to allow the two steps to proceed almost on--shell,
which means that the energy defect should be small. At the
threshold for $\eta$ production $\Delta E_0\approx -15$~MeV, which
illustrates the \emph{kinematic miracle} first noted for the
\pdhe\ reaction~\cite{Kilian}. We then neglect the dependence of
the individual $pp\to\pi^+d$ and $\pi^+n\to\eta p$ amplitudes upon
the Fermi momentum so that the only place where \bmath{q} occurs
is in the denominator of the propagator. By defining a Lorentz
transformed velocity with components
\begin{equation}
\bmath{V}'_{\bot}=\bmath{V}_{\bot}\ \ \textrm{and}\ \
\bmath{V}'_{\|}=\gamma\bmath{V}_{\|}
\end{equation}
the whole integrand can be written in terms of $\bmath{q}'$.

After decomposing the deuteron wave function into its $\ell=0,\,2$
components, we are left with integrals of the form
\begin{eqnarray}
\nonumber%
&&i\int\rd^3q'\frac{\tilde{\varphi}_{\ell}(q')T_{\ell}(\hat{\bmath{q}'})}
{\Delta E_0+\bmath{V}'\cdot\bmath{q}'+i\varepsilon} =
\int_0^{\infty} \rd t\,e^{i(\Delta E_0+i\varepsilon)t}\\
\nonumber%
&&\times\int_0^{\infty}q'^2\,\rd q'\varphi_{\ell}(q')\,
\int\rd\Omega_{q'}\,e^{it\bmath{V}'\cdot\bmath{q}'}\,
T_{\ell}(\hat{\bmath{q}'})\\
&&= S_{\ell}(\Delta E_0,|\bmath{V}'|)\,
T_{\ell}(\hat{\bmath{V}'})\:,
\end{eqnarray}
where the $S$-- and $D$--state form factors involve integrals over
the \emph{configuration--space} deuteron wave functions
\begin{equation}
S_{\ell}(\Delta
E_0,|\bmath{V}'|)=\frac{(2\pi)^{3/2}}{|\bmath{V}'|}\,
\int_0^{\infty}\rd t\, e^{i\omega t}\,\varphi_{\ell}(t)
\end{equation}
with $\omega = \Delta E_0/|\bmath{V}'|$.

The $s$--wave nature of the $S_{11}(1535)$ resonance that
dominates low energy $\eta$ production in $\pi^+n\to\eta\,p$,
leads to largely isotropic production at low energies. We can then
take the matrix element to be proportional to the corresponding
spin--non--flip amplitude
\begin{equation}
\mathcal{M}(\pi^+n\to\eta\,p)= \frac{4\pi\sqrt{s_{\eta
p}}}{m}\,f(\pi^+n\to\eta\,p)\:.
\end{equation}

For simplicity, of the six invariant $pp\to \pi^+d$ amplitudes we
shall keep only the largest in our energy region, for which
\begin{equation}
\mathcal{M}(pp\to \pi^+d)=-\sqrt{2}\,\mathcal{A}\,
(\bmath{\epsilon}_d^{\dagger}\cdot\hat{\bmath{p}}_{\pi})\,
\phi_{pp}\:,
\end{equation}
where $\phi_{pp}$ and $\bmath{\epsilon}_d$ represent the
spin--zero and spin--one initial and final $NN$ states. This is
then related to the cm cross section through
\begin{equation}
\left|\mathcal{A}\,\right|^2=8(2\pi)^2\left\{\frac{s_{pp}}{m^2}
\frac{p_p}{p_{\pi}}\,\frac{\rd\sigma}{\rd\Omega}(pp\to\pi^+d)\right\}_{\!\!cm}.
\end{equation}

In addition to the $\pi^+$ diagram of fig.~1c, there is also the
possibility of $\pi^0$ propagation between the two steps. This can
be taken into account by simply multiplying the matrix element
$\mathcal{M}$ of eq.~(\ref{master}) by an isospin factor of $3/2$.
Using eq.~(\ref{PSG2}), we then arrive at an expression for the
distribution of the \pde\ cross section in terms of the final
deuteron angles and $\eta p$ invariant mass:
\begin{eqnarray}
\nonumber%
&&\frac{\rd^2\sigma}{\rd\Omega_d\,\rd m_{\eta p}}
=\frac{9}{4(2\pi)^4}\, \frac{s_{\eta p}\,
m_d\,p_d\,p_{\eta}^*}{s\,p\,E_{\pi}^2}
\left(\left|S_S\right|^2+\left|S_D\right|^2\right)\times\\
\nonumber%
&&\left\{\frac{s_{pp}}{m^2}
\frac{p_p}{p_{\pi}}\,\frac{\rd\sigma}{\rd\Omega}(pp\to\pi^+d)\right\}_{\!\!cm}
\left\{\frac{p_{\pi}}{p_{\eta}}\sigma_{\textrm{\tiny
tot}}(\pi^+n\to\eta\,p)\right\}_{\!\!cm},\\
\end{eqnarray}
where we have taken advantage of the isotropy of the
$\pi^+n\to\eta\,p$ differential cross section to integrate over
the $\eta$ angles and to write the result in terms of the total
cross section. This option is not open for the other differential
distributions, where extra non--trivial integrations have to be
performed.

An effective range description of the low--energy
$\pi^+n\to\eta\,p$ total cross section is
\begin{equation}
\frac{p_{\pi}^*}{p_{\eta}^*} \sigma_{\textrm{\tiny
tot}}(\pi^+n\to\eta\,p) \approx
\frac{2.76|a|^2}{\left|1-iap_{\eta}^*+\half
r_0ap_{\eta}^{*2}\right|^2}\,.
\end{equation}
The data base has not improved significantly since we took the
values $a=(0.476+0.279i)\,$fm and
$r_0=(-3.16-0.13i)\,$fm~\cite{FW}.

The $pp\to \pi^+d$ cross sections are obtainable from the SAID
analysis~\cite{SAID} but, because the intermediate pion in
Fig.~\ref{fig1}a is non--physical, some prescription is needed to
extrapolate from the experimental data. We have assumed in the
applications that the amplitudes are unchanged when the cm
production angle is kept fixed.

%
%
\section{Comparison with experiment}
\setcounter{equation}{0}%
\vspace{-1cm}

\input epsf
\begin{figure}[htb]
\begin{center}
\mbox{\epsfxsize=3.5in \epsfbox{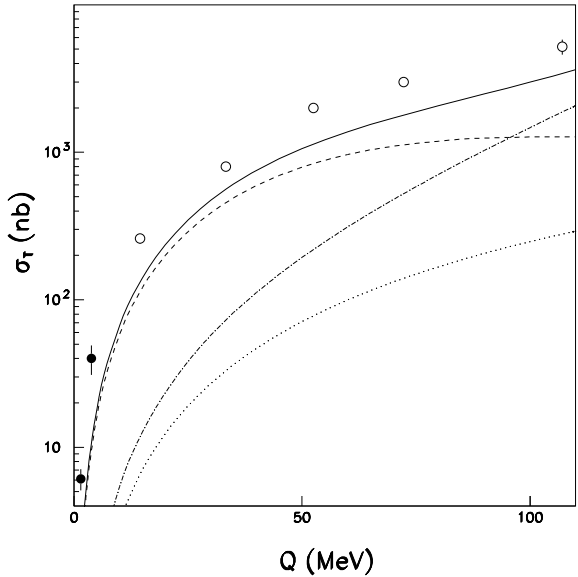}} \caption{Total cross
section for the reaction \pde. The experimental data are taken
from Saclay (solid circles)~\cite{Hibou} and CELSIUS (open
circles)~\cite{Jozef2}. The dotted curve represents the
predictions of the triangle diagram (impulse approximation), the
chain curve the pick--up contribution, and the dashed curve that
of the two--step model. The incoherent sum of these three terms is
shown by the solid curve. \label{fig2}}
\end{center}
\end{figure}
\vspace{-0.5cm}

The predictions of the three driving terms of Fig.~\ref{fig1} are
compared in Fig.~\ref{fig2} with the existing experimental
data~\cite{Hibou,Jozef2}. It is immediately apparent that the
impulse approximation of the triangle graph in Fig.~\ref{fig1}b
lies about two orders of magnitude below the experimental data.
The ambiguity in the relative phase of the amplitudes
$\mathcal{W}_{1,s}$ and $\mathcal{W}_{0,t}$ is therefore largely
irrelevant and the impulse approximation term will now be dropped
from further discussion.

In the near--threshold region of low $Q$, the two-step
contribution dominates that of the pick--up by an order of
magnitude. The kinematics here are somewhat similar to those of
low--energy \pdhe\ reaction where the pick--up term with a
proton--deuteron final--state interaction similarly underpredicts
the experimental data~\cite{Laget,Germond}. However, the two--step
term levels off at $Q\approx 80$~MeV and eventually decreases at
higher energies due to a combination of factors. The region where
the intermediate pion is close to being physical becomes a smaller
fraction of the allowed phase space but also the $pp\to d\pi^+$
cross section falls steeply above the $\Delta$ region. Since the
pick--up term must approach that for quasi--free $pn\to d\eta$
above the $NN$ threshold, the two contributions must cross before
then. In our estimates, this occurs at $Q\approx 95$~MeV.

The incoherent sum of the three contributions is also shown in
Fig.~\ref{fig2}. Though the shape is very similar to that of the
experimental data, these lie about a factor of two above the curve
at high energies and a little bit more near threshold. This
parallels a similar underprediction of the \pdhe\ reaction in a
two--step model that includes the effects of the $\eta\,^3$He
\fsi~\cite{FW}. The fractions of phase space where the $pd$ or
$\eta d$ final--state interactions might enhance the $p\,d\,\eta$
channel are relatively small at the higher energies.

\input epsf
\begin{figure}[htb]
\begin{center}
\mbox{\epsfxsize=3.8in \epsfbox{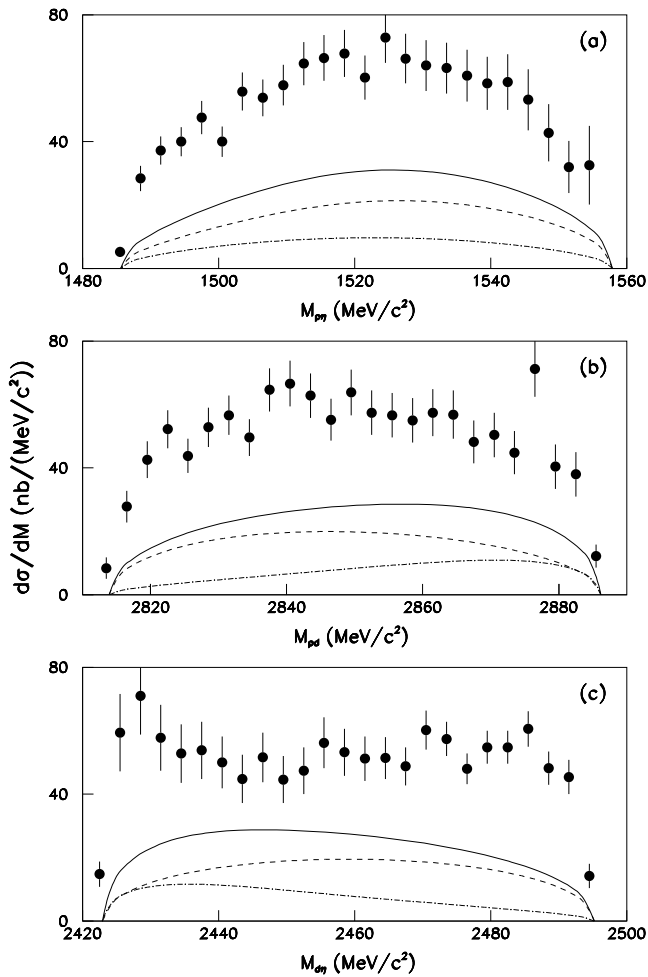}}
\caption{Distributions in the invariant masses of the \pde\
reaction at $Q=72.3$~MeV. The panels (a), (b), and (c) refer
respectively to the $p\eta$, $pd$, and $d\eta$ cases. The
experimental data from Ref.~\cite{Jozef2} are compared to the
predictions of the pick--up term (chain curve), the two--step
model (dashed curve), and their incoherent sum (solid curve).
\label{fig3}}
\end{center}
\end{figure}

More information can be derived from the differential
distributions that were measured at $T_p=1032$~MeV
($Q=72.3$~MeV)~\cite{Jozef2}. The only significant enhancement in
the three invariant mass distributions of Fig.~\ref{fig3} is that
near the $\eta d$ threshold, which is consistent with the large
scattering length suggested by the low energy $pn\to d\eta$ total
cross section data~\cite{Stina}. In contrast, despite the
existence of the nearby $^3$He bound--state pole, there is no sign
of any $pd$ \fsi. The two--step model leads to only minor
distortions of the invariant mass phase spaces, pushing
$m_{p\eta}$ to slightly larger values, whereas the pick--up
contribution preferentially populates low $m_{d\eta}$ masses which
is reflected kinematically as a slight increase at higher values
of $m_{pd}$.

\input epsf
\begin{figure}[htb]
\begin{center}
\mbox{\epsfxsize=3.5in \epsfbox{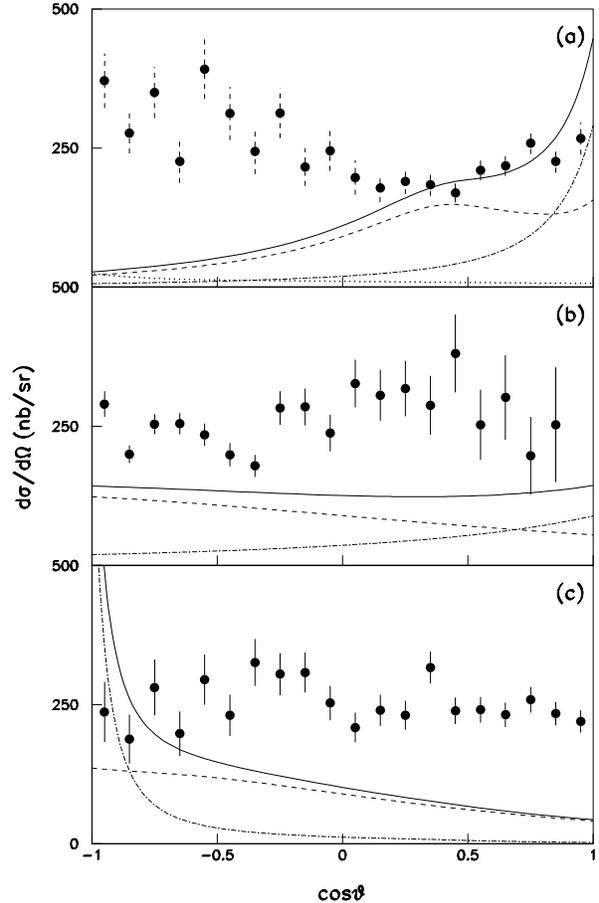}}
\caption{Centre--of--mass angular distributions with respect to
the incident proton direction of the \pde\ reaction at
$Q=72.3$~MeV. The panels (a), (b), and (c) refer respectively to
outgoing $d$, $\eta$, and $p$. The experimental data from
Ref.~\cite{Jozef2} are compared to the predictions of the pick--up
term (chain curve), the two--step model (dashed curve), and their
incoherent sum (solid curve). \label{fig4}}
\end{center}
\end{figure}

Apart from the overall strength being too low by a factor of two,
the sums of the two contributions give plausible descriptions of
the invariant mass distributions but the same cannot be said for
the angular distributions shown in Fig.~\ref{fig4}. In the
pick--up contribution of Fig.~\ref{fig1}a, the final proton is a
spectator and the transverse momentum is governed by the Fermi
momentum components in the deuteron. This automatically gives a
sharp peak for a cm proton angle close to the backward direction
and this has a slightly weaker kinematic reflection around the
forward deuteron direction. Given that the estimate within the
pick--up model has relatively few uncertainties, the discrepancy
with the experimental data of Ref.~\cite{Jozef2} is particularly
significant.

There are far more uncertainties in the evaluation of the
two--step model since the intermediate pion is generally off its
mass shell and this effects the kinematics of the $pp\to d\pi^+$
reaction. However, the deuteron angular distribution of
Fig.~\ref{fig4}a does suggest that the models need to be
supplemented at large angles and this might be the reason for the
underestimation of the total cross section.
%
%
\section{Conclusions}
\setcounter{equation}{0}%

We have estimated the contributions of three different models to
the total and partial cross sections of the \pde\ reaction below
the $\eta$--production threshold in nucleon--nucleon collisions.
The impulse approximation turns out to be largely negligible
compared to the other two terms of which the two--step model is
shown to be dominant in the near--threshold region. However,
although some of our kinematic approximations may start to break
down before one reaches the free $pn$ threshold ($Q\approx
192$~MeV), the estimates suggest that the two--step mechanism
would provide only a very small correction to the pick--up
interpretation of the CELSIUS quasi--free $pn\to d\eta$
data~\cite{Stina}.

We have neglected the final--state interactions which should
distort the lower edges of the invariant mass distributions shown
in Fig.~\ref{fig3} while having a smaller effect on the total
cross section away from the threshold region. In fact the only
\fsi\ clearly seen in the experimental data is that associated
with the $\eta d$ channel~\cite{Jozef2}. Now the total cross
section at the higher energies is underestimated by a factor of
two, which is a very similar factor to that found for the
near--threshold two--body \pdhe\ reaction interpreted in the same
two--step approach, after the inclusion of the necessary \fsi\
effects~\cite{FW}. It is, however, intriguing to note that,
although there is no sign of the strong $pd$ \fsi\ in the
experimental data of Fig.~\ref{fig3}b, the normalisation of the
total \pde\ total cross section is predicted successfully from the
\pdhe\ data using the final--state extrapolation
theorem~\cite{extrapolation} without implementing a full dynamical
model.

Apart from the \fsi\ regions, it is hard to draw firm conclusions
on the models from the invariant mass distributions of
Fig.~\ref{fig3}. There is far more information to be gleaned from
the angular distributions of Fig.~\ref{fig4}. The models
underpredict the data in the backward deuteron hemisphere and,
more critically, there is no sign in the data for the sharp peaks
for forward--going deuterons and backward--going protons being
produced in the pick--up process. Though these might be softened
somewhat by multiple scatterings or \fsi, it should also be noted
that if either the deuteron \underline{or} the proton is lost down
the beam pipe then the event is not registered~\cite{Jozef2}.

If indeed the two--step model is the dominant mechanism well below
the nucleon--nucleon threshold, it would have consequences for the
$p\,d \to ppn\eta$ four--body final state, where one could expect
to see a strong $pn$ \fsi. Data on this reaction were taken at
CELSIUS simultaneously with those on the other channels, but their
detailed analysis is not yet complete~\cite{Jozef3}. Data also
exist from COSY on the analogous sub--threshold $p\,d\to
K^+\Lambda\, d$ reaction~\cite{Valdau} and it would seem likely
that the corresponding two--step model should play an important
role there as well.
%
%
\begin{acknowledgement}
We are much indebted to our colleagues in Uppsala, especially
J.~Z\l oma\'nczuk, for discussions regarding the
$\eta$--production programme at CELSIUS. One of the authors (CW)
is grateful for financial support and hospitality from both the
Department of Radiation Sciences and the The Svedberg Laboratory
of the University of Uppsala. Comments by Y.~Uzikov regarding the
impulse approximation term proved helpful.
\end{acknowledgement}
%
%

%
%
\end{document}